\documentclass[aps,twocolumn,showpacs]{revtex4}%
\usepackage{amsfonts}
\usepackage{amsmath}
\usepackage{amssymb}
\usepackage{graphicx}%
\begin{document}
\title{Optical M\"{o}bius Singularities }
\author{Isaac Freund}
\affiliation{Physics Department, Bar-Ilan University, Ramat-Gan ISL52900, Israel}

\begin{abstract}
M\"{o}bius strips with one, two, three, and four, half-twists are shown to be
generic features of three-dimensional (nonparaxial) elliptically polarized
light. \ The geometry and topology of these unusual singularities is
described and the multitude of winding numbers that characterize their
structures is enumerated; probabilities for the appearance of different
configurations are presented.

\end{abstract}


\maketitle

The singularities of optical fields are major determinants of the field
structure that are of intense current interest [$1-13$]. \ Here we extend
previous studies of polarization singularities in three dimensional (3D)
optical fields [$14-22$], finding new, highly unusual structures.

3D light, i.e. light composed of waves with a broad spread of noncoplanar
propagation directions, is almost invariably elliptically polarized. \ The
generic singularities of 3D optical (and other electromagnetic) ellipse fields
are lines of circular polarization, C lines, and lines of linear polarization,
L lines [$19-22$]. \ At the point where a C line (L line) pierces a plane,
$\Sigma$, a C point (L point) appears. \ The principal axes of the
polarization ellipse are the major axis, the minor axis, and the ellipse
normal. \ At a C point the ellipse degenerates into a circle, the C circle,
and the major and minor axes are undefined (singular). \ At an L point the
ellipse collapses to a line, the major axis of the ellipse, and the minor axis
of, and the normal to, the ellipse become undefined.

All three principal axes of the ellipses surrounding the singularity, however,
remain well defined. The projections onto $\Sigma$ of the major and minor axes
(minor axis and normal) of those ellipses whose centers lie in this plane
rotate about the C point (L point) with characteristic winding number $\pm1/2$
($\pm1$) [$19-22$]. \ Throughout this report, for C points (L points) we take
$\Sigma$ to be the plane of the C circle (to be oriented perpendicular to the
major axis). \  We call this plane the \emph{principal plane}, and label it
$\Sigma_{0}$.%

\begin{figure}
[h]
\includegraphics[width=0.3\textwidth]%
{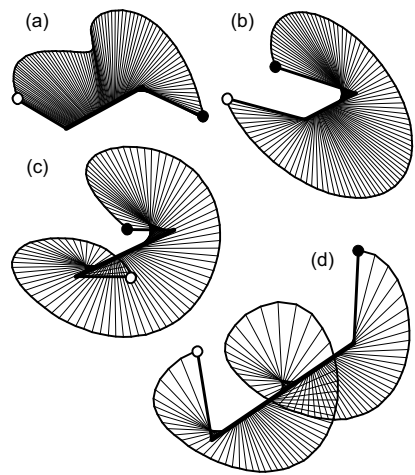}%
\caption{Optical M\"{o}bius singularities in a random ellipse field. \ For
clarity, the M\"{o}bius strips formed by the polarization ellipses whose
centers lie on a small circle in $\Sigma_{0\text{ }}$are opened and
straightened. \ Ellipse axes are shown by the thin straight lines that spiral
around the thick straight line formed by the ellipse centers; only the upper
half of each axis is shown. The small white (black) circle marks the ellipse
at the arbitrary \textquotedblleft beginning\textquotedblright%
\ (\textquotedblleft end\textquotedblright) of the opened M\"{o}bius strip;
when the strip is closed these two ellipses are adjacent. \ \ For an odd
(even) number of half twists the white and black circles are on opposite sides
(the same side) of the line of ellipse centers. (a)-(d) M\"{o}bius strips
formed by the major axes (a)-(c) or ellipse normals (d) of the polarization
ellipses surrounding: (a) a C point - this (wrinkled) strip has a single half
twist; (b) an ordinary (nonsingular) point - this strip has one full twist;
(c) a C point that is different from the one in (a) - this strip has three
half twists; (d) an L point - this strip has two full twists. \ The M\"{o}bius
strip in (a) forms a right-handed screw and has topological index $-1/2$, the
strips in (b) - (d) form left-handed screws and have indices $+1$, $+3/2$, and
$+2$, respectively. \ In random fields right/left-handed strips occur with
equal probabilities. \ Although a circular path is chosen here for simplicity,
any simple path that encloses only the C or L point yields (a possibly highly
wrinkled) M\"{o}bius strip with the same twist index as does the circle}%
\label{FigMS1}%
\end{figure}

Previous studies of 3D light have concentrated on projections onto the
principal plane $\Sigma_{0}$ of the polarization ellipses whose centers lie in
this plane; here we study the \emph{full} 3D arrangement of these ellipses.
\ Choosing for simplicity concentric circular paths in $\Sigma_{0}$ that
surround a C or L point, we find that for $\sim80\%$ of all C points both the
major and minor axes of the ellipses on each path generate a M\"{o}bius strip
with a single half twist, whereas for the remaining $20\%$ the M\"{o}bius
strip has three half twists. \ M\"{o}bius strips on paths surrounding L points
are generated both by the minor axes of, and the normals to, the surrounding
ellipses, and have two full twists. \ Previously [$14$], we showed that the
polarization ellipses surrounding ordinary (nonsingular) points generate
M\"{o}bius strips with one full twist, so that together with the results
reported here, the polarization ellipses in 3D electromagnetic fields
generically produce M\"{o}bius strips with one half-, two half-, three half-,
and four half-twists. \ Representative M\"{o}bius strips are shown in Fig.
\ref{FigMS1}.

Simple physical models of M\"{o}bius strips with the same topology as those in
Fig. \ref{FigMS1} can be constructed from short thin rods and a long strip of
somewhat flexible material. \ For a model of a C point (L point) the rods
represent say the major axis (ellipse normal). \ The \textquotedblleft
upper\textquotedblright\ half of each rod is painted black, the rods are
oriented parallel to one another and attached at their centers to the flexible
strip such that they are perpendicular to the long axis of the strip.
\ Introducing the appropriate number of half twists generates a model of the
desired figure in Fig. \ref{FigMS1}; gluing the ends of the strip together
completes the construction of a model of the M\"{o}bius strip that surrounds a
C or an L point.

By continuity, for C points (L points) the angle that say the major axis
(ellipse normal) of the surrounding ellipses makes with the principal plane
goes to zero as one approaches the central singularity. \ In the physical
model described above this can be accomplished simply by squashing the
M\"{o}bius strip flat. \ Such a deformation changes the geometry of the strip
but preserves its topology; the deformations in real optical ellipse fields
are, of course, rather more subtle.

In optical ellipse fields there are no packing problems due to overlap of the
axes of neighboring ellipses in the concentric, nested M\"{o}bius strips
surrounding C and L points. \ The reason is that the polarization ellipse
describes the time variation of the oscillating electric field at a point, so
that the ellipse has dimension zero, whereas the M\"{o}bius strip has
dimension one, and is a line.

The M\"{o}bius strips in Fig. \ref{FigMS1} were measured in a simulated 3D
random optical field (speckle pattern), where the optical field, $\mathbf{E}$,
in complex representation, was generated as a sum of $100$ plane waves with
random phases and randomly oriented transverse linear polarizations. \ Many
waves, however, are not required for C and L lines and their M\"{o}bius
strips; all these features are found in three beam optical lattices.

The C and L lines used in Fig. \ref{FigMS1} were obtained from the above
simulation by tracking zeros of the C and L point discriminants $D_{C}%
=a_{1}^{2}-4a_{2}$, and $D_{L}=a_{2}$; these discriminants are obtained from
the characteristic equation $\lambda^{2}+a_{1}\lambda+a_{2}=0$ of the real
coherency matrix $M_{ij}=\operatorname{Re}(E_{i}^{\ast}E_{j});i,j=x,y,z$
[$16$]. The orientation of the major axis $\boldsymbol{\alpha}$ and the minor
axis $\boldsymbol{\beta}$ of the ellipses surrounding a C or L point were
calculated using formulas due to Berry [$15,18$]: $\boldsymbol{\alpha
}=\operatorname{Re}(\mathbf{E}^{\ast}\sqrt{\mathbf{E}\boldsymbol{\cdot
}\mathbf{E}})$, $\boldsymbol{\beta}=\operatorname{Im}(\mathbf{E}^{\ast}%
\sqrt{\mathbf{E}\boldsymbol{\cdot}\mathbf{E}})$; the normal to the ellipse,
$\boldsymbol{\gamma}$, was calculated using $\boldsymbol{\gamma}%
=\operatorname{Im}\left(  \mathbf{E}^{\ast}\mathbf{\times E}\right)  $ [$19$].

\begin{figure}
[h]
\includegraphics[width=0.35\textwidth]%
{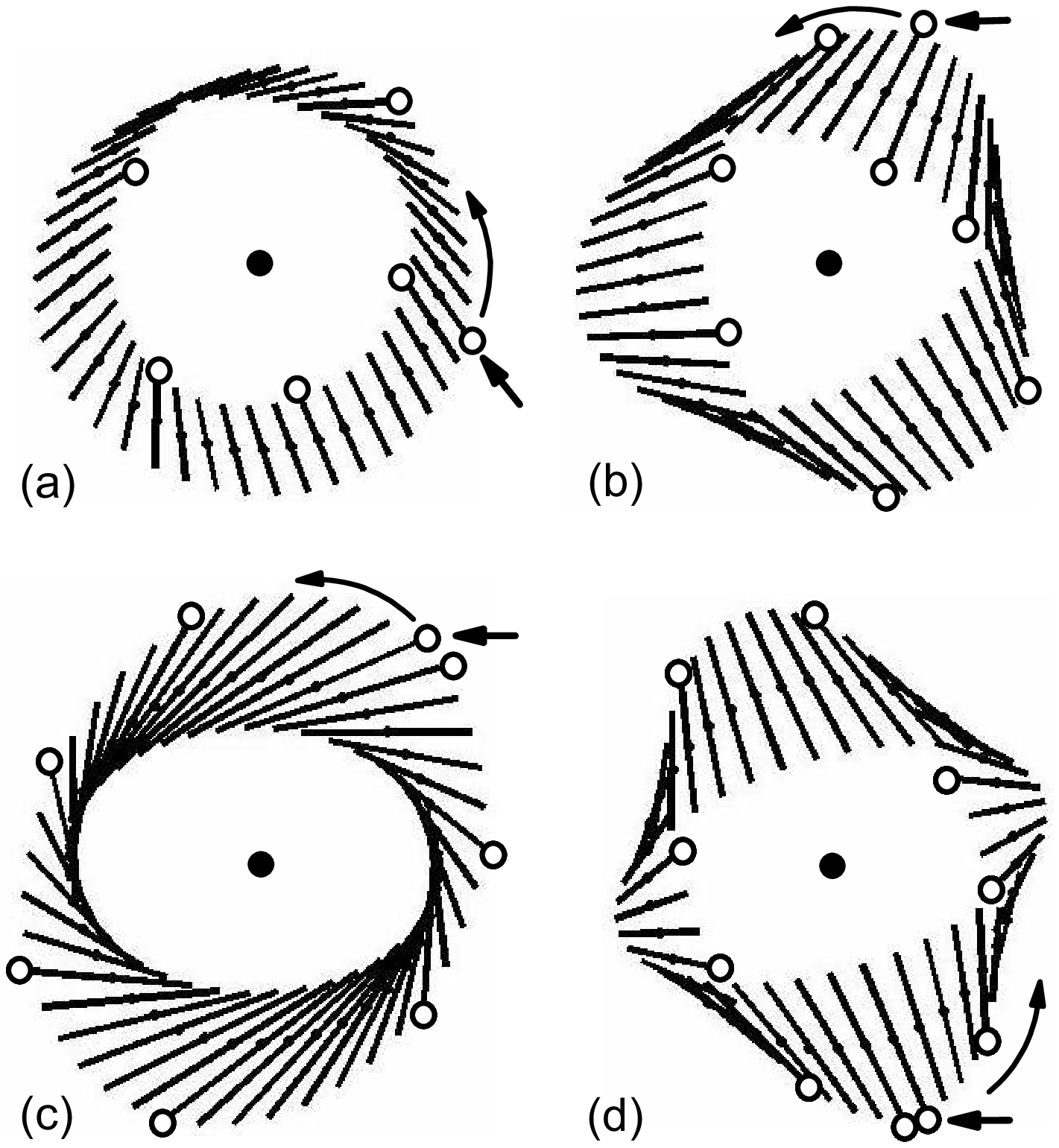}%
\caption{C point indices $I_{\alpha}^{\left(  C\right)  }$ and $I_{\gamma
}^{\left(  C\right)  }$. \ $I_{\alpha}^{\left(  C\right)  }$ ($I_{\gamma
}^{\left(  C\right)  }$) is generated by the projections onto the principal
plane $\Sigma_{0\text{ }}$of axis $\boldsymbol{\alpha}$ (axis
$\boldsymbol{\gamma}$)\ of the ellipses surrounding the C point.
\ Conventionally, the centers of the surrounding ellipses are taken to be
located on a small circle surrounding the C point, although any simple closed
path produces the same index. \ Projections of the axes of the surrounding
ellipses onto $\Sigma_{0\text{ }}$are shown by short straight lines, ellipse
centers by small black dots, and the C point by the central black circle. \ As
an aid in following the rotation of the axes one end is decorated by a small
white circle. \ The here counterclockwise direction in which the surrounding
circle (path) is traversed is shown by a curved arrow, the starting point of
the path by a small straight arrow. \ (a) $I_{\alpha}^{\left(  C\right)  }$
for the C point in Fig. \ref{FigMS1}(a). \ The net rotation of the ellipse
axes is in the same counterclockwise direction as the path, so the sign of
$I_{\alpha}^{\left(  C\right)  }$ is positive. \ The axes rotate by $180^{o}$
during one $360^{o}$ traverse of the path, so the magnitude of the winding
number is $1/2$. \ Adding in the sign, yields $I_{\alpha}^{\left(  C\right)
}=+1/2$. \ (b) $I_{\alpha}^{\left(  C\right)  }$ for the C point in Fig.
\ref{FigMS1}(c). The ellipses rotate by $180^{o}$ in the clockwise
(retrograde) direction and $I_{\alpha}^{\left(  C\right)  }=-1/2$. \ (c)
$I_{\gamma}^{\left(  C\right)  }$ for the C point in Fig. \ref{FigMS1}(a).
\ The ellipses rotate by $360^{o}$ in the same clockwise direction as the path
and $I_{\gamma}^{\left(  C\right)  }=+1$. \ $I_{\gamma}^{\left(  C\right)
}=+1$ also for the C point in Fig. \ref{FigMS1}(c). \ (d) $I_{\gamma}^{\left(
C\right)  }$ for another ellipse located on the same C line as the ellipse in
Fig. \ref{FigMS1}(c). \ Here $I_{\gamma}^{\left(  C\right)  }=-1$. }%
\label{FigIndexIac}%
\end{figure}

The three primary topological indices that characterize a C point in a 3D
field are: (\emph{i}) The well known [$19$] winding number $I_{\alpha}%
^{(C)}\equiv I_{\beta}^{\left(  C\right)  }=\pm1/2$ of the projection onto the
principal plane $\Sigma_{0}$ of the major or minor axes of the surrounding
ellipses; (\emph{ii}) The winding number $I_{\gamma}^{(C)}=\pm1$ (not
discussed previously) of the projection onto $\Sigma_{0}$ of the normals to
the surrounding ellipses; and (\emph{iii}) the first of the two major findings
of this report: the twist index $\tau_{\alpha}^{\left(  C\right)  }=$
$\tau_{\beta}^{\left(  C\right)  }=\pm1/2,\pm3/2$ of the M\"{o}bius strips
generated by the major and minor axes of the surrounding ellipses. \ For C
points, axis $\mathbf{\gamma}$ does not generate a M\"{o}bius strip. $\ $The
above three indices generate $2\times2\times4=16$ different C points, an
eight-fold enlargement of previous C point characterizations [$19$]. \ All
$16$ C points appear in our simulations; $\tau_{\alpha}^{\left(  C\right)  }$
is illustrated in Fig. \ref{FigMS1}, $I_{\alpha}^{\left(  C\right)  }$ and
$I_{\gamma}^{\left(  C\right)  }$ are illustrated in Fig. \ref{FigIndexIac}.%

The complex field components $E_{x}$, $E_{y}$, and $E_{z}$, in the immediate
vicinity of any point in a 3D field can be expanded to first order (the order
required here) as $E_{m}=K_{m}+\sum_{n=x,y,z}\left(  P_{mn}+iQ_{mn}\right)
u_{n},$ where $m=x,y,z,$ and $u_{x}=x$, $u_{y}=y$, $u_{z}=z$. \ Without loss
of generality, for C points we take $K_{x}$ to be real, set $K_{y}%
=i\sigma\left\vert K_{x}\right\vert $, where $\sigma=\pm1$, and set $K_{z}=0$
and $u_{z}=z=0$. \ This describes a C point at the origin with the normal to
the C circle oriented along the $z$-axis, and the surrounding ellipses lying
in the principal plane $\Sigma_{0}$ of the C point, here the $xy$-plane.
\ Similarly, for L points we take $K_{z}$ to be real, set $K_{x}=K_{y}=0$ and
$u_{z}=z=0$. \ This describes an L point at the origin with its major axis
oriented along the $z$-axis, and again places the surrounding ellipses in the
$xy$-plane, the principal plane $\Sigma_{0}$ of the L point. \ In both cases,
in the principal plane $\Sigma_{0}$, the generally nonzero parameters $P_{nz}$
and $Q_{nz}$ make no contribution because their multiplier, $z$, is zero. \ The field component $E_{z}$ is not zero, however, because it involves terms in $x$ and $y$ that are not zero. \ We note that all structures found in our simulations were reproduced using this expansion with suitable choices for the expansion parameters $K,P,Q$.
\ Analytical expressions can be obtained for the various indices in terms of
these expansion parameters by extending the methods of [$14$]; these methods
require a discussion that is too lengthy to be included here.

Close facsimiles of the measured random field strips in Fig. \ref{FigMS1} can
be generated from simpler expansions. \ The M\"{o}bius strip in Fig.
\ref{FigMS1}(a), for example, is reasonably well approximated by
$Ex=-5+y\left(  i-1\right)  $, $Ey=-5i+x$, $Ez=-ix+y$; that in Fig.
\ref{FigMS1}(b) by $Ex=-5$, $Ey=i$, $Ez=-y$; the one in Fig. \ref{FigMS1}(c)
by $Ex=10-y$, $Ey=10i-x$, $Ez=y$; and the strip in Fig. \ref{FigMS1}(d) by
$Ex=i\left(  2y-x\right)  $, $Ey=x+iy$, $Ez=5$.

Preliminary statistics of the various topological indices were obtained by
choosing random values for the nonzero $K$, the various $P$ and $Q$, and the
sign of $\sigma$. \ A random number generator that produced a normal
distribution with unit variance was used. \ The data listed below were
obtained from $1.6\times10^{6}$ independent samples.

\begin{figure}
[h]
\includegraphics[width=0.475\textwidth]%
{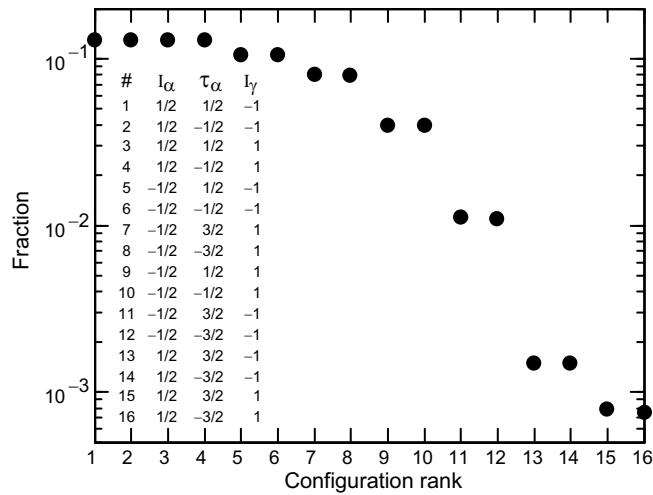}%
\caption{C point Zipf plot.\ Configuration fraction (probability) vs.
configuration rank for configurations ranked by decreasing probabilities.}%
\label{FigZipfC}%
\end{figure}

Fig. \ref{FigZipfC} presents a modified Zipf plot for the three C point
indices $I_{\alpha}^{(C)}$, $I_{\gamma}^{(C)}$, and $\tau_{\alpha}^{\left(
C\right)  }$.%

There are four primary topological indices that characterize an L point. \ The
first two indices are the well known [$19$] winding number $I_{\beta}^{\left(
L\right)  }\equiv I_{\gamma}^{\left(  L\right)  }=\pm1$, and the index
$I_{\alpha}^{\left(  L\right)  }=\pm1$ (not discussed previously). \ These
indices are analogous to the corresponding C point indices $I_{\alpha
}^{\left(  C\right)  }\equiv I_{\beta}^{\left(  C\right)  }$, and $I_{\gamma
}^{\left(  C\right)  }$. \ The other two indices are the second major finding
of this report: the two independent twist indices, $\tau_{\beta}^{\left(
L\right)  }=\pm2$, and $\tau_{\gamma}^{\left(  L\right)  }=\pm2$. \ The
M\"{o}bius strip in Fig. \ref{FigMS1}(d) was generated by axis $\mathbf{\gamma
}$ and has index $\tau_{\gamma}^{\left(  L\right)  }=+2$, whereas the strip
generated by axis $\mathbf{\beta}$ for the same L point has index $\tau
_{\beta}^{\left(  L\right)  }=-2$. \ For L points, axis $\mathbf{\alpha}$ does
not generate a M\"{o}bius strip.

\begin{figure}
[h]
\includegraphics[width=0.475\textwidth]%
{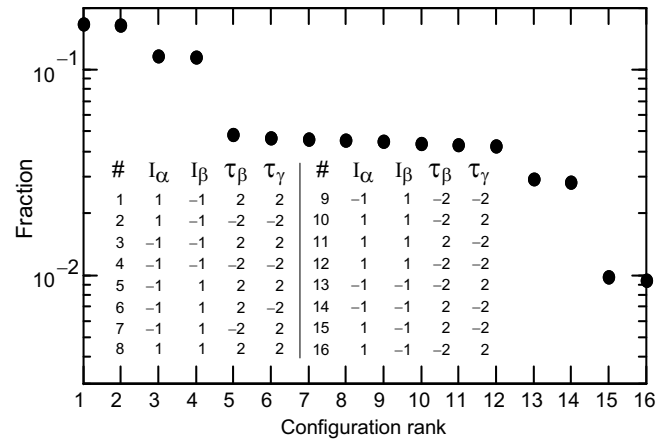}%
\caption{L point Zipf plot.}%
\label{FigZipfL}%
\end{figure}

These four L point indices generate $2^{4}=16$ different L points, all of
which appear in our simulations. \ Fig. \ref{FigZipfL} presents a modified
Zipf plot that summarizes the different L point configurations and their
relative probabilities.%

\begin{figure}
[h]
\includegraphics[width=0.45\textwidth]%
{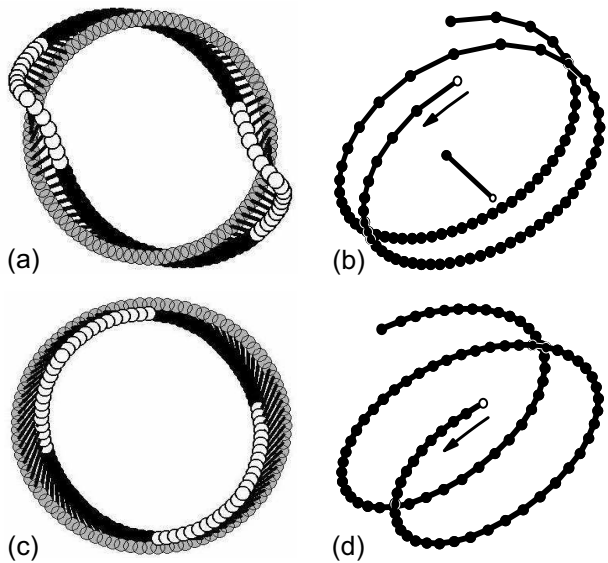}%
\caption{L point structures. \ The ellipses in $\Sigma_{0}$ that surround an L
point can generate one of two geometrically distinct structures: (a),(b) a
double-full-twist M\"{o}bius strip, or (c),(d) a section of a cone. \ Ellipse
half-axes of the surrounding ellipses are shown by thick straight lines, the
centers of these ellipses by light gray circles;the L point itself (not shown)
lies at the center of this circle. \ \ The end points of the ellipse axes are
shown by filled white (black) circles if these endpoints lie above (below) the
plane of ellipse centers. In (a) the ellipse endpoints wind around the circle
of ellipse centers to form a pair of interlocking rings. \ This M\"{o}bius
strip is the one shown in Fig. 1(d). \ As shown here in (b), an observer
walking along the circle of ellipse centers sees the endpoints spiral around
her path (central straight line); the winding number of this spiral is
$\tau=+2$. \ In (c) the endpoint circle and the circle of ellipse centers do
not interlock, and this structure is not a M\"{o}bius strip. An observing
walking along the circle of ellipse centers sees a section of an inverted cone
whose "bottom" edge is her path and whose "top" edge, formed by the ellipse
endpoints, is fluted and deeply scalloped; these endpoints form the spiral in
(d), which has winding number $\tau=+2$. \ \ Inverted cones, such as the one
shown here, and cones in which the ellipse axes flare outward, instead of
inward, from the circle of ellipse centers are equally probable, as are
positive ($+2$) and negative ($-2$) values for $\tau$. }%
\label{FigMS2}%
\end{figure}

There are four geometrically distinct types of L points. \ We label these
L$_{mm}$, L$_{mc}$, L$_{cm}$, and L$_{cc}$, where the first subscript refers
to the figure generated by axis $\mathbf{\beta}$ of the surrounding ellipses,
the second to axis $\mathbf{\gamma}$; $m$ ($c$) implies that the figure is a
M\"{o}bius strip (\textquotedblleft cone\textquotedblright). \ The differences
between Mobius strips and cones are shown in Fig. \ref{FigMS2}. \ Although it
is relatively easy to distinguish visually between Mobius strips and cones, it
is difficult to do so with complete reliability using a computer program.
\ From visual examination of some $100$ L points we observed that L$_{mm}$
points appear mostly, but not exclusively, for negative $I_{\beta}^{\left(
L\right)  }$ and L$_{cc}$ points appear mostly, but not exclusively, for
positive $I_{\beta}^{\left(  L\right)  }$; L$_{mc}$ and L$_{cm}$ points appear
less frequently than do L$_{mm}$ and L$_{cc}$ points. \ The data in Fig.
\ref{FigZipfL} do not distinguish between topological equivalent but
geometrically distinct L points: this figure lists all L points with a given
index configuration without regard to whether these points are L$_{mm}$,
L$_{mc}$, L$_{cm}$, or L$_{cc}$.%

The topological indices discussed here are all of first order in the sense
that the field modulations that give rise to them grow linearly with distance
from the singularity; these indices should be observable in microwave
[$3,4,20$] and in interferometric, phase-sensitive, nanoprobe optical
[$23-27$] experiments that are able to probe the three-dimensional structure
of the field.

C and L lines and their associated C and L points correspond to degeneracies
of symmetric second order tensors (matrices) [$17,18$]. \ Accordingly, they,
and their associated M\"{o}bius strips, can be expected to occur, and should
therefore be sought, in other physical line and vector fields described by
such tensors; candidates include liquid crystals, strain fields, and flow fields.

I thank Prof. David A. Kessler for helpful comments and suggestions.

\begin{center}
\textbf{------------------------}
\end{center}

\hspace*{-0.14in}[1] R. I. Egorov, M. S. Soskin, D. A. Kessler, and I. Freund,
Phys. Rev. Lett. \textbf{100}, 103901 (2008).

\hspace*{-0.14in}[2] K. O'Holleran, M. R. Dennis, F. Flossmann, and M. J.
Padgett, Phys. Rev. Lett. \textbf{100}, 053902 (2008).

\hspace*{-0.14in}[3] S. Zhang and A. Z. Genack, Phys. Rev. Lett. \textbf{99},
203901 (2007)

\hspace*{-0.14in}[4] S. Zhang, B. Hu, P. Sebbah, and A. Z. Genack, Phys. Rev.
Lett. \textbf{99}, 063902 (2007).

\hspace*{-0.14in}[5] R. Pugatch, M. Shuker, O. Firstenberg, A. Ron, and N.
Davidson, Phys. Rev. Lett. \textbf{98}, 203601 (2007).

\hspace*{-0.14in}[6] O. Peleg, G. Bartal, B. Freedman, O. Manela, M. Segev,
and D. N. Christodoulides, Phys. Rev. Lett. \textbf{98}, 103901 (2007).

\hspace*{-0.14in}[7] W. Wang, Z. Duan, S. G. Hanson, Y. Miyamoto, and M.
Takeda, Phys. Rev. Lett. \textbf{96}, 073902 (2006).

\hspace*{-0.14in}[8] Y. F. Chen, T. H. Lu, and K. F. Huang, Phys. Rev. Lett.
\textbf{96}, 033901 (2006).

\hspace*{-0.14in}[9] F. Flossmann, U. T. Schwarz, M. Maier, and M. R. Dennis,
Phys. Rev. Lett. \textbf{95}, 253901 (2005).

\hspace*{-0.14in}[10] H. F. Schouten, T. D. Visser, G. Gbur, D. Lenstra, and
H. Blok, Phys. Rev. Lett. \textbf{93}, 173901 (2004).

\hspace*{-0.14in}[11] G. A. Swartzlander, Jr. and J. Schmit, Phys. Rev. Lett.
\textbf{93}, 093901 (2004).

\hspace*{-0.14in}[12] D. M. Palacios, I. D. Maleev, A. S. Marathay, and G. A.
Swartzlander, Jr., Phys. Rev. Lett. \textbf{92}, 143905 (2004).

\hspace*{-0.14in}[13] D. R. Solli, C. F. McCormick, R. Y. Chiao, S. Popescu,
and J. M. Hickmann, Phys. Rev. Lett. \textbf{92}, 043601 (2004).

\hspace*{-0.14in}[14] I. Freund, Opt. Commun. \textbf{256}, 220 (2005); ibid
\textbf{249}, 7 (2005).

\hspace*{-0.14in}[15] M. V. Berry, J. Opt. A \textbf{6}, 675 (2004).

\hspace*{-0.14in}[16] I. Freund, J. Opt. A \textbf{6}, S229 (2004).

\hspace*{-0.14in}[17] M. V. Berry and M. R. Dennis, Proc. Roy. Soc. Lond. A
\textbf{457}, 141 (2001).

\hspace*{-0.14in}[18] M. V. Berry in \emph{Singular Optics 2000}, ed. M. S.
Soskin, Proc. SPIE \textbf{4403}, 1 (2001).

\hspace*{-0.14in}[19] J. F. Nye, \emph{Natural Focusing and Fine Structure of
Light} (IOP Publ., Bristol, 1999).

\hspace*{-0.14in}[20] J. V. Hajnal, Proc. Roy. Soc. Lond. A \textbf{430}, 413
(1990); ibid \textbf{447} (1987); ibid \textbf{414}, 433 (1987).

\hspace*{-0.14in}[21] J. F. Nye and J. V. Hajnal, Proc. Roy. Soc. Lond. A
\textbf{409}, 21 (1987).

\hspace*{-0.14in}[22] J. F. Nye, Proc. Roy. Soc. Lond. A \textbf{389}, 279
(1983); ibid \textbf{387}, 105 (1983).

\hspace*{-0.14in}[23] Z. H. Kim and S. R. Leone, Opt. Express \textbf{16},
1733 (2008).

\hspace*{-0.14in}[24] K. G. Lee, H. W. Kihm, J. E. Kihm, W. J. Choi, H. Kim,
C. Ropers, D. J. Park, Y. C. Yoon, S. B. Choi, D. H. Woo, J. Kim, B. Lee, Q.
H. Park, C. Lienau, D. S. Kim, Nature Photon. \textbf{1}, 53 (2006).

\hspace*{-0.14in}[25] C. Rockstuhl, I. Marki, T, Scharf, M. Salt, H. P.
Herzig, and R. Dandliker, Current Nanoscience \textbf{2}, 337 (2006).

\hspace*{-0.14in}[26] R. Dandliker, I. Marki, M. Salt, and A. Nesci, J. Opt.
A. \textbf{6}, S189 (2004).

\hspace*{-0.14in}[27] M. L. M. Balistreri, H. Gersen, J. P. Korterik, L.
Kuipers, and N. F. van Hulst, Science \textbf{294}, 1080 (2001).

\end{document}